\documentclass[aps,prl,twocolumn,superscriptaddress, preprintnumbers,nofootinbib]{revtex4-1}

\usepackage{epsf,epsfig,amsmath,amssymb,amsfonts}
\usepackage{color}
\usepackage{slashed}

\newsavebox{\ns}
\newsavebox{\dbrane}
\renewcommand{\arraystretch}{1.2}
\def\be{\begin{equation}}
\def\ee{\end{equation}}
\def\bea{\begin{eqnarray}}
\def\eea{\end{eqnarray}}

\def\Dslash{\,\,{\raise.15ex\hbox{/}\mkern-12mu D}}
\def\Dbarslash{\,\,{\raise.15ex\hbox{/}\mkern-12mu {\bar D}}}
\def\delslash{\,\,{\raise.15ex\hbox{/}\mkern-9mu \partial}}
\def\delbarslash{\,\,{\raise.15ex\hbox{/}\mkern-9mu {\bar\partial}}}
\def\pslash{\,\,{\raise.15ex\hbox{/}\mkern-9mu p}}
\def\calDslash{\,\,{\raise.15ex\hbox{/}\mkern-12mu {\cal D}}}

\newcommand\R{\mathbb{R}}

\newcommand\T{\mathbb{T}}
\newcommand\diff{\mbox{d}}

\newcommand{\vol}{\mbox{vol}}

\newcommand{\nn}{\nonumber \\}

\newcommand{\dd}{\diff}
\newcommand{\DD}{\textrm{D}}

\def\T{\Theta}

\allowdisplaybreaks

\begin{document}

\title{Warped Ricci-flat reductions}

\preprint{DMUS-MP-14/07}

\preprint{YITP-SB-1418}

\preprint{CAS-KITPC/ITP-437}

 \author{E. \'O Colg\'ain}
 \affiliation{C.N.Yang Institute for Theoretical Physics, SUNY Stony Brook, NY 11794-3840,USA \& \\ Department of Mathematics, University of Surrey, Guildford GU2 7XH, UK}
 \author{M.M. Sheikh-Jabbari}
 \affiliation{School of Physics, Institute for Research in Fundamental Sciences (IPM),  P.O.Box 19395-5531, Tehran, Iran}\affiliation{Department of Physics, Kyung Hee University, Seoul 130-701, Korea}
 \author{J.F. V\'azquez-Poritz}
 \affiliation{Physics Department
New York City College of Technology, The City University of New York 300 Jay Street, Brooklyn NY 11201, USA \& \\ The Graduate School and University Center, The City University of New York 365 Fifth Avenue, New York NY 10016, USA}
 \affiliation{Kavli Institute for Theoretical Physics China, Institute of Theoretical Physics, Chinese Academy of Sciences, Beijing 100190, China}
\author{H. Yavartanoo}
\affiliation{State Key Laboratory of Theoretical Physics, Institute of Theoretical Physics,  Chinese Academy of Sciences, Beijing 100190, China}
\author{Z. Zhang}
\affiliation{Physics Department
New York City College of Technology, The City University of New York 300 Jay Street, Brooklyn NY 11201, USA \& \\ The Graduate School and University Center, The City University of New York 365 Fifth Avenue, New York NY 10016, USA}

\begin{abstract}
We present a simple class of warped-product vacuum (Ricci-flat) solutions to ten and eleven-dimensional supergravity, where the internal space is flat and the warp factor supports de Sitter (dS) and anti-de Sitter (AdS) vacua in addition to trivial Minkowski vacua. We outline the construction of consistent Kaluza-Klein (KK) reductions and show that, although our vacuum solutions are non-supersymmetric, these are closely related to the bosonic part of well-known maximally supersymmetric reductions on spheres. We comment on the stability of our solutions, noting that (A)dS$_3$ vacua pass routine stability tests.
\end{abstract}

\maketitle

\setcounter{equation}{0}

\section{Introduction} \label{Introduction}

Studying gravity in various dimensions has had many different motivations, at the classical and quantum levels, dating back to the 1920's and the seminal works of Kaluza and Klein. String theory, of course, prefers ten or eleven-dimensional supergravity theories and related compactifications or reductions to lower dimensions.
Except for rather limited though important cases in lower dimensions, such as supersymmetric solutions to ungauged supergravity \cite{Gibbons:1982fy, Gauntlett:2002nw},
we are not even close to classifying all solutions to a given (super)gravity theory.  In the absence of supersymmetry, a specific class of simpler solutions that is often studied in gravity theories involves ``vacuum'' solutions, i. e. solutions to Einstein equations with vanishing energy-momentum tensor, or alternatively, Ricci-flat geometries.

AdS/CFT motivations have directed a lot of activity in gravity solution construction towards finding and classifying solutions that involve an AdS factor and an internal compact space. Such solutions are almost always not vacuum solutions and involve various form-field fluxes present in supergravity theories. Moreover,  for AdS/CFT purposes and also for stability requirements, as well as having (quantum) corrections under control, it is often demanded that such solutions preserve a fraction of the global supersymmetry of the theory. This tendency has led to Ricci-flat solutions with AdS factors being largely overlooked.

Separately, solutions with a de Sitter factor have also been of particular interest within higher-dimensional supergravity and string theory setups,  as the observed universe we are living in seems to be an asymptotically de Sitter space. Nonetheless, it has proven to be notoriously difficult to construct four-dimensional de Sitter solutions in a string theory setting which are classically and quantum mechanically stable and do not have a moduli problem \cite{Burgess:2003ic, Caviezel:2008tf}. A leading framework \cite{KKLT} and its uplifting procedure to produce dS vacua has recently been called into question \cite{Bena:2009xk}, leading to a renewed interest in alternatives \cite{Danielsson:2013rza}. In this paper, we consider simple higher-dimensional gravity with no local sources or non-perturbative contributions and simply solve the equations of motion.  As a result, the de Sitter vacua we find are some of the simplest in the literature and, while we forfeit supersymmetry from the onset, we still retain some control through a scaling limit.

Although our AdS solutions are singular - albeit in a ``good" sense \cite{Gubser:2000nd} - it is a striking feature of our de Sitter constructions that they are completely smooth. This should be contrasted with recent studies of persistent singularities in non-compact geometries where anti-branes are used to uplift AdS vacua \cite{Bena:2012ek} (see also \cite{Junghans:2014wda}). Here, since we are considering vacuum solutions, we have no branes, and thus, no singularities. Any branes that do exist only appear when we turn on fluxes in the lower-dimensional theories to construct dS$_3$ vacua. Regardless of whether fluxes are turned on or not, our construction evades the well-known ``no-go" theorem \cite{Maldacena:2000mw} on the basis that the internal space is non-compact. 

The scaling limit we employ may be traced to ``near-horizon" limits of Extremal Vanishing Horizon (EVH) black holes \cite{Hossein-4d-EVH}. Within an AdS/CFT context, such warped-product solutions
have been studied previously in \cite{Fareghbal:2008ar, Fareghbal:2008eh} (more recently \cite{More-AdS5-EVH, DeWolfe:2013uba}), where consistent KK reductions to lower-dimensional theories were constructed. In contrast to usual AdS/CFT setups, the vacua of the lower-dimensional theories are supported exclusively through the warp factor and lift to vacuum solutions in ten and eleven dimensions. We show that if one neglects the possibility of a large number of internal dimensions,  there are just three example in this class. Furthermore, we demonstrate that the solutions can be easily generalized to de Sitter and that consistent KK reductions exist as scaling limits of well-known sphere reductions, for example \cite{Cvetic:1999xp}.

We begin in the next section by considering a general $D$-dimensional warped-product spacetime Ansatz on the assumption that the internal space is Ricci-flat. While it is most natural to consider $\mathbb{R}^q$, the same analysis holds for Calabi-Yau cones and we expect a variant to hold for more generic Calabi-Yau. Locally, this construction encompasses cases for which the internal space is an Einstein space such as a sphere, which is conformally flat and the conformal factor is automatically included by way of the warp factor. We identify a class of vacuum solutions where the internal Ricci scalar can be made to vanish by tuning the dimensions. Somewhat surprisingly, this leads to three isolated examples, which only reside in ten or eleven dimensions,  supporting (A)dS$_{p}$ vacua with $p=5, 6, 8$. We remark that supersymmetry is broken.

In the following section, we focus on the warped (A)dS$_5$ solution to eleven-dimensional supergravity, where the internal space is $\mathbb{R}^6$. We explicitly construct a KK reduction Ansatz and note that the lower-dimensional theory one gets is five-dimensional U(1)$^3$ gauged supergravity \cite{Cvetic:1999xp} on the nose. Importantly, this observation guarantees stability within our truncation, though of course instabilities can arise from modes that are truncated out \cite{Bobev:2010ib}. In the absence of warping, the (A)dS$_5$ vacuum becomes  a Minkowski vacuum and one can compactify the internal space to get the usual KK reduction to \textit{ungauged} supergravity on a Calabi-Yau manifold. We show that the KK reduction naturally arises as a scaling limit of the KK reduction of eleven-dimensional supergravity on S$^7$, further truncated to the Cartan U(1)$^4$ \cite{Cvetic:1999xp}. Using a similar scaling argument, we also exhibit KK reductions from ten and eleven-dimensional supergravity on $\mathbb{R}^4$ and $\mathbb{R}^3$, respectively.

Since none of our solutions are manifestly supersymmetric, the final part of our paper  concerns (classical) stability. In constructing the KK reductions, we have assumed a product structure for the internal space and it is a well-known fact that reductions based on product spaces are prone to instabilities where the volume of one subspace increases whilst another decreases \cite{Duff:1984sv,Berkooz:1998qp,{DeWolfe:2001nz}}. Neglecting all solutions to the five-dimensional theory, which are guaranteed to be stable within the truncation, we find that the AdS$_6$ and AdS$_8$ vacua are unstable. By constructing lower-dimensional Freund-Rubin type solutions within our AdS$_p ,\ p=5,6,8$ truncations, we show that AdS$_3$ solutions are stable. On physical grounds, since these arise as the near-horizon of EVH black holes \cite{Fareghbal:2008ar,Fareghbal:2008eh}, they are expected to be classically stable.

In the final part of this paper, we construct an example of a dS$_3$ vacuum and study its stability.  We show that the vacuum energy of the de Sitter solution can be tuned so as to stabilise the vacuum against tunneling. This guarantees that the vacuum would be suitably long lived and, along with other dS$_3$ solutions to string theory \cite{Dong:2010pm}. We close the paper with some discussion of related open directions.

\section{Ricci-flat solutions}
\label{sec:Ricci}
In this section we identify a class of Ricci-flat solutions in general dimension $D = p+q$. From the offset, we assume that the overall spacetime takes the form of a warped product,
\be
\dd s^2_{p+q} = \Delta^{m}\ \dd s^2(M_p) + \Delta^n \ \dd s^2(\Sigma_q),
\ee
decomposed into a $p$-dimensional external spacetime $M_p$ and a $q$-dimensional Ricci-flat internal space $\Sigma_q$. $m, n$ denote constant exponents and the warp factor $\Delta$ only depends on the coordinates of the internal space.

Denoting external coordinates, $a, b= 0,\dots, p-1$ and internal coordinates $m, n = 1, \dots q$, the vanishing of the internal Ricci scalar, i. e. $g^{mn} R_{mn}$, yields the equation:
\bea
\label{Eq0}
[\frac{mp}{2} &+& n (q-1) ] \nabla^2 \Delta = \biggl[ n (q-1) - \frac{mp}{2} (m-1) \nn &-& (q-2) [\frac{n^2}{4} (q-1) + \frac{mnp}{4}] \biggr] \Delta^{-1} (\partial \Delta)^2.
\eea

If the internal space is $\mathbb{R}^q$ and $r$ denotes its radial coordinate, then simple solutions to (\ref{Eq0}) are given by
\be
\Delta = \biggl\{ \begin{array}{c} (c_1 +  c_2 r^{2-q})^{\frac{1}{(1+\kappa)}}, ~~q \neq 2,  \\
(c_1 + c_2 \log r )^{\frac{1}{(1+ \kappa)}}, ~~q=2,
\end{array}
\ee
where $\kappa$ is a constant that depends on $m, n, p, q$ and $c_i$ denote integration constants. Geometries with these $\Delta$ are singular at the origin $r=0$.

One may hope to find non-singular solutions to \eqref{Eq0},  by forcing the bracketed terms  to vanish:
\be
\label{condition:mnpq}
m = 2 - \frac{4}{q}, \quad n = - \frac{4}{q}, \quad p = \frac{4(q-1)}{q-2}.
\ee
This condition defines our class of Ricci-flat solutions. Note that $q=2$ is not a legitimate choice and if one demands integer dimensions, we are hence led to only the following choices for $(p, q$): ($5, 6$), ($6,4$) and ($8,3$). Besides these choices, one can also formally consider large $D$, $q \rightarrow \infty$ limit where one encounters a four-dimensional vacuum ($p=4$).

It is a curious property of this class of solutions that they only exist in ten and eleven (and also infinite) dimensions, settings where we have low-energy effective descriptions for string theory. From the onset, there is nothing outwardly special about our Ansatz and one would assume that examples could be found for general $D$, yet we find that this is not the case.

To specify the overall spacetime, we simply now have to record the warp factor. Again, evoking the existence of a radial direction, we can write $\Delta$ as
\be
\label{warp_factor}
\Delta = \sqrt{1+ \lambda r^2},
\ee
where we have normalized the integration constants, one of which, $\lambda = -1$ or $+1$, dictates whether the vacuum is anti-de Sitter or de Sitter spacetime, respectively. For $\lambda=0$, the warp factor becomes trivial and the solution reduces to $D$-dimensional Minkowski spacetime. The radius, $\ell$, of $M_p$ is expressed as
\be\label{radius-ell}
\ell^2 = \frac{1}{|\lambda|} \frac{(p-1)}{(q-2)}.
\ee

The other $D$-dimensional vacuum Einstein equations yield the following equation for  $\Delta$:
\be
\Delta \nabla^2 \Delta + (\partial \Delta)^2 = q \lambda,
\ee
and, as a result, it is easy to infer through
\be
\int_{\Sigma_q} \left( \Delta \nabla^2 \Delta + (\partial \Delta)^2 -q \lambda\right) = - q \lambda \vol(X) = 0,
\ee
that a compact internal space $\Sigma_q$ requires a Minkowski vacuum, $\lambda = 0$. Therefore, all our (A)dS spacetimes,  will have non-compact internal spaces.

Despite the existence of a covariantly constant spinor (which is a result of Ricci-flatness), it is easy to see that none of these geometries are supersymmetric, except when $\lambda = 0$. As a cross-check, we note that where complete
classifications of supersymmetric solutions exist, for example \cite{Gauntlett:2004zh}, one can confirm that our solutions are not among them.

As an extension to the $\Sigma_q=\mathbb{R}^q$ internal space,  for $q=4$ and $6$,  $\Sigma_q$ can easily be chosen to be a Calabi-Yau cone over a Sasaki-Einstein space, or more generally by a cone over an Einstein space. However, such cones have a conical singularity at their apex.

We also remark that for $\lambda =-1$, we encounter a curvature singularity at $r=1$, where the warp factor vanishes. This does not affect the Ricci scalar, since the above solutions are Ricci-flat, but it does show up in contractions of the Riemann tensor $R_{MNPQ} R^{MNPQ}$. This singularity can be seen to be of ``good" type \cite{Gubser:2000nd}, a point that was made recently in \cite{DeWolfe:2013uba} and we will return to in due course.  In contrast, the Minkowski and de Sitter solutions are smooth.

\section{KK Reduction on the solutions}

Taking each of the warped-product solutions identified in the previous section in turn, one can construct simple consistent Kaluza-Klein reductions to the lower-dimensional theory. By ``simple", in contrast to traditional reductions, we mean that there is a clear division between scalars in the metric and gauge fields in the fluxes of the higher-dimensional supergravity. This means that even with scalars the overall spacetime is Ricci-flat\footnote{Superficially, this bears some similarity to the AdS/Ricci-flat correspondence \cite{Caldarelli:2012hy} in that there is a connection between a Ricci-flat space and an AdS spacetime. However, our connection, which also works at the level of the equations of motion, does not involve an analytic continuation of dimensionality.}, whereas the inclusion of gauge fields leads to a back-reaction externally, with the internal space remaining Ricci-flat. In this section we discuss the KK reduction of each of the three solutions independently.

\subsection{KK Reduction on the $\mathbb{R}^6$}

We start with a reduction from eleven-dimensional supergravity to five dimensions on $\mathbb{R}^6$. The main idea for such a reduction was presented in
\cite{Fareghbal:2008eh}.
Our warped Ansatz naturally generalises the known reduction to a Minkowski vacuum on Calabi-Yau, for example \cite{Elvang:2004ds}. Since the lower-dimensional theory in that case is \textit{ungauged} supergravity, here we present a simple Ansatz that recovers the bosonic sector of $D=5$ U(1)$^3$ gauged supergravity \cite{Cvetic:1999xp}, and via a flip in a sign (appropriate double Wick rotation and analytic continuation), the AdS$_5$ vacuum becomes dS$_5$. We recall the five-dimensional action of $D=5$ U(1)$^3$ gauged supergravity:
\bea
\label{U13theory}
\mathcal{L}_5 &=& R * \mathbf{1} - \frac{1}{2} \sum_{i}^2 \dd \varphi_i \wedge * \dd \varphi_i  - \frac{1}{2} \sum_i^3 X_i^{-2} F^i \wedge * F^i \nn
&-& 4 \lambda \sum_i^3 X_i^{-1} \vol_5
+ F^1 \wedge F^2 \wedge A^3.
\eea
In the above action, $F^i = \dd A^i$ and the scalars $X_i$ are subject to the constraint $\prod_{i=1}^3 X_i = 1$. In terms of the unconstrained scalars $\varphi_i$, they may be expressed as
\be
\label{x}
X_1 = e^{-\frac{1}{2} \left( \frac{2}{\sqrt{6}} \varphi_1 + \sqrt{2} \varphi_2 \right) }, \quad X_2 = e^{-\frac{1}{2} \left( \frac{2}{\sqrt{6}} \varphi_1 - \sqrt{2} \varphi_2 \right) },
\ee
where $X_3 = (X_1 X_2)^{-1}$. It is a well-known fact that the theory (\ref{U13theory}), with coupling $g^2 = - \lambda$, arises from a KK reduction of type IIB supergravity on S$^5$ \cite{Cvetic:1999xp}. Here, we provide an alternative higher-dimensional guise.

The U(1)$^3$ theory (\ref{U13theory}) arises from the following KK reduction Ansatz:
\bea
\label{redansatz}
\dd s_{11}^2 &=& \Delta^{\frac{4}{3}} \dd s^2(M_5) + \Delta^{-\frac{2}{3}} \sum_{i=1}^3 X_i \left( \dd \mu_i^2 + \mu_i^2 \dd \psi_i^2 \right), \nn
G_4 &=& - \sum_i \mu_i \dd \mu_i \wedge \dd \psi_i \wedge \dd A^i,
\eea
where the internal space is a product of three copies of $\mathbb{R}^2$ and the warp factor is given in (\ref{warp_factor}) in terms of the overall radius $r$, where $r^2 = (\mu_1^2 + \mu_2^2 + \mu_3^2)$. The four-form is largely self-selecting, since it is only this combination of external and internal forms that will scale in the same way with the Ricci tensor, $R_{ab} \sim \Delta^{-\frac{4}{3}} \bar{R}_{ab} $.  In general, constructing KK reductions for warped product spacetimes is tricky and a useful rule of thumb is that the fluxes should scale in the same way as the Ricci tensor, so that the warp factor drops out of the Einstein equation. Further discussion can be found in \cite{OColgain:2011ng}.

The internal part of metric in the reduction Ansatz \eqref{redansatz}  is rather unusual in the sense that we have isolated U(1)'s in the internal metric but have not gauged the isometries.
In fact, it can be shown that the inclusion of traditional KK vectors along the U(1)'s in the metric would result in terms that scale differently with the warp factor. Therefore, on their own, they are inconsistent but it may be possible to restore consistency, essentially by mixing the metric with the fluxes so that the overall factor that appears with the gauge fields scales correctly. This would involve a more complicated Ansatz-- potentially one where Ricci-flatness is sacrificed-- and we leave this to future work.

This reduction is performed at the level of the equations of motion and is, by definition, consistent. The flux equation of motion in eleven-dimensional supergravity, $\dd *_{11} G_4 + \frac{1}{2} G_4 \wedge G_4 = 0$, leads to the lower-dimensional flux equations of motion, while Ricci-flatness along each copy of $\mathbb{R}^2$ leads to an equation of motion for $X_i$. The constraint on the $X_i$ comes from cross-terms in the Ricci tensor of the form
\be
R_{a \mu_i} = - \mu_i \Delta^{-\frac{7}{3}} X_i^{-\frac{1}{2}} \partial_{a} \log \prod_i X_i = 0.
\ee
The reduction proves to be inconsistent at the level of the action. This is probably due to the non-compactness of the internal space  \footnote{It was also observed in \cite{Itsios:2012dc} that a non-compact reduction was inconsistent when performed at the level of the action.} .

Some further comments are now in order. As stated, one may readily show that the $D=11$ uplift of the AdS$_5$ vacuum is not supersymmetric. \textit{A priori}, there is nothing to rule out the possibility that the solutions to U(1)$^3$ gauged supergravity, which are supported by the scalars $X_i$ and the gauge fields $A_i$,  are supersymmetric. However, we believe that this is unlikely. To back this up, we have  confirmed that a three-parameter family of wrapped brane solutions considered in \cite{Benini:2013cda} (see \cite{Cucu:2003bm} for earlier works) is not supersymmetric in the current context. The same solutions are supersymmetric when uplifted on $S^5$.

As for compactness, for $\lambda =-1$ we have a natural cut-off on the internal spaces, namely $ \sum_{i=1}^3 \mu_i^2 \leq 1$, thus leading to a finite-volume internal space, where each $\mathbb{R}^2$ subspace may be regarded as a disk. Curvature singularities of this type have been identified as the ``good" type in the literature \cite{Gubser:2000nd} and their CFT interpretation has been explored in \cite{Fareghbal:2008eh}. For $\lambda=+1$ this is a smooth embedding of $D=5$ U(1)$^3$ de Sitter gravity in eleven-dimensional supergravity, albeit with a non-compact internal space.

\subsection{Origin of the KK reduction}
It is striking that we have arrived at a class of vacua that only reside in ten and eleven-dimensional supergravity. In this section,  we offer an explanation as to why that may be the case. Our observation hinges on a known ``far from BPS" near-horizon limit of certain extremal black holes in $D=4$ U(1)$^4$ and $D=5$ U(1)$^3$ gauged supergravity \cite{Fareghbal:2008ar,Fareghbal:2008eh} (see also \cite{DeWolfe:2013uba}), where an AdS$_3$ near-horizon is formed by incorporating an internal circular direction with the scalars, in this case $X_i$, all scaled appropriately.

Eschewing explicit solutions, we are free to apply the same scaling for the scalars $X_i$ directly to the KK reduction Ansatz presented in \cite{Cvetic:1999xp}. For concreteness, we consider the SO($8$) KK reduction on S$^7$, further truncated to the U(1)$^4$ Cartan subgroup. The reduction Ansatz may be written as \cite{Cvetic:1999xp}
\bea
\dd s^2_{11} &=& \Delta^{\frac{2}{3}} \dd s^2_4 + \Delta^{-\frac{1}{3}} \sum_{i=1}^4 X_i^{-1} \left[ \dd \mu_i^2 + \mu_i^2 \DD \phi_i^2 \right], \\
G_{4} &=& 2 \sum_i \biggl[ \left( X_i^2 \mu_i^2 - \Delta X_i \right) \vol_4 +  \frac{1}{2} *_4 \dd \log X_i \wedge \dd (\mu_i^2) \nn
&-& \frac{1}{2} X_i^{-2} \dd (\mu_i^2) \wedge \DD \phi_i \wedge *_4 F^i \biggr] ,
\eea
where we have defined $\DD = \dd \phi_i + A^i$, $\Delta = \sum_{i=1}^4 X_i \mu_i^2$, $F^i = \dd A^i$ and $\mu_i$ are constrained by $\sum_{i=1}^4 \mu_i^2 =1$. The $X_i$ are subject to the constraint $\prod_{i=1}^4 X_i=1$.

We now isolate $X_1$  and blow it up by taking the limit\footnote{In terms of the unconstrained scalars, $\varphi_i, i=1, 2, 3$, this simply corresponds to the limit $\varphi_i \rightarrow - \infty$, so it is symmetric.}
\bea
X_1&=& \epsilon^{-\frac{3}{2}} \tilde{X}_1, \quad \quad X_i = \epsilon^{\frac{1}{2}} \tilde{X}_i, ~~i = 2, 3, 4,  \nn \phi_1 &=& \epsilon^{-1} \varphi_1, \quad~~~~  g_{4} = \epsilon \tilde{g}_4, 
\eea
with $\epsilon \rightarrow 0$. In the process, the internal $\phi_1$ direction migrates and combines with the original four-dimensional metric to form a five-dimensional subspace. Performing this scaling at the level of the Ansatz yields
\bea
\dd s^2_{11} &=& \mu_1^{\frac{2}{3}} \left[  \tilde{X}_1^{\frac{2}{3}} \dd \tilde{s}^2_4 + \tilde{X_1}^{-\frac{4}{3}} \dd \varphi_1^2 \right] \nn &+&  \mu_1^{-\frac{2}{3}} \sum_{i=2}^4 \tilde{X_1}^{-\frac{1}{3}} \tilde{X}_i^{-1} \left[ \dd \mu_i^2 + \mu_i^2 \dd \phi_i ^2 \right], \\
G_{4} &=& -\frac{1}{2} \sum^4_{i=2} \tilde{X}_i^{-2} \dd (\mu_i^2) \wedge \dd \phi_i  \wedge *_4 F^i,
\eea
where $\mu_1$ is constrained, so the warp factor is
\be
\mu_1 = \sqrt{1 - (\mu_2^2 + \mu_3^2 + \mu_4^2)}.
\ee
Up to redefinitions, and the introduction of $\lambda$, this Ansatz is the same as the consistent KK reduction Ansatz identified in the previous section. Note that the $*_4 F^i$ in the original notation refers to a two-form, with the Hodge dual of the two-form leading to a two-form in the new five-dimensional spacetime, which appears wedged with the volume of the internal disks. It is interesting that the AdS vacuum is now sourced by the warp factor and not by the original $\vol_4$ term in the four-form flux, $G_4$, which is suppressed in the limiting procedure.

Note that this limiting procedure naturally leads to a singularity, which conforms to the ``good" type under the criterion of \cite{Gubser:2000nd}, since the scalar potential is bounded above in the lower-dimensional potential. In other words, from the perspective of the original $D=4$ U(1)$^4$ gauged supergravity, the limiting procedure results in a steadily more negative potential.

One could first take a limit where the S$^7$ degenerates to S$^5\times \R^2$ or S$^3\times \R^4$ \cite{Cvetic:1999pu} and then apply the scaling limit of this paper. The result of this two-step process is that the warp factor does not depend on the $\R^2$ or $\R^4$ portion of the internal space that resulted from taking the first limit. One could then perform dimensional reduction and T-duality along these flat directions so that our KK reductions are reinterpreted as arising from scaling limits of S$^5$ or S$^3$ reductions of type IIB theory, along with a toroidal reduction along the remaining flat directions. The resulting lower-dimensional theory admits a domain wall, rather than (A)dS, as a vacuum solution.

\subsection{KK reductions on $\mathbb{R}^4$}
In this subsection, we briefly record the other consistent KK reductions with warp factors. We begin by considering type IIB supergravity
 on the product $\mathbb{R}^4 \equiv \mathbb{R}^2\times \mathbb{R}^2$. It was previously shown in \cite{Fareghbal:2008ar} that there is a consistent KK reduction to a six-dimensional theory admitting an AdS$_6$ vacuum with a single scalar and three-form flux. We recall the ten-dimensional Ansatz \footnote{Our $H_3$ is related to $F_3$ in \cite{Fareghbal:2008ar} by a factor of two, $F_3 = 2 H_3$.} but allow for the opposite sign in the warp factor,
\bea
\dd s_{10}^2 &=& \Delta \dd s^2(M_6) + \Delta^{-1} \sum_{i=1}^2 L^2 e^{Y_i} (\dd \mu_i^2 + \mu_i^2 \dd \psi_i^2), \nn
F_5 &=& (1+ *_{10}) H_3 \wedge \mu_1 \dd \mu_1 \wedge \dd \psi_1,
\eea
where $\Delta = \sqrt{1  +\lambda (\mu_1^2 + \mu_2^2)}$, we have relabeled $X_i = e^{Y_i}$ to avoid logarithms and $L$ is a length scale. The Bianchi identity for $F_5$ implies that we can define a two-form potential so that $H_3 = \dd B_2$ and the self-duality requirement dictates that both $H_3$ and its Hodge dual appear in the Ansatz for the five-form flux.  In \cite{Fareghbal:2008ar}, it was found in the absence of a dilaton and axion that consistency of the reduction (considering cross-terms in the Ricci tensor and flatness condition) requires $Y_1 = -Y_2$, leaving a single scalar in six dimensions. We recall that for the SO(6) reduction of type IIB supergravity on S$^5$ the dilaton and axion do not feature in the scalar potential (see for example \cite{Cvetic:2000nc}), so it is expected that one can reinstate them here.

After a conformal transformation to get to the Einstein frame, the six-dimensional action takes the form
\bea
\label{D6_theory}
\mathcal{L}_6 &=& \sqrt{-\hat{g}_6} \biggl(\hat{R}_6 - \frac{1}{2} (\partial \Phi)^2 - \frac{1}{2} e^{2 \Phi} (\partial \chi)^2 - \frac{1}{4} (\partial Z)^2 \nn  &-& \frac{8}{L^2} \lambda \cosh \frac{Z}{2} - \frac{1}{12} e^{-Z} H_3^2 \biggr),
\eea
where $\Phi = Y_1 + Y_2$ and $Z=Y_1 - Y_2$.

It is worth noting that the three-form entering in the lower-dimensional action is not self-dual from the six-dimensional perspective but it is self-dual in the five-form flux of type IIB supergravity. Truncating out either $Y_1$ or $Y_2$, we arrive at the action of \cite{Fareghbal:2008ar}, up to a symmetry of that action $Z \leftrightarrow Z^{-1}$.

Since the internal space is flat, the (A)dS$_6$ vacuum of the six-dimensional reduced theory could be embedded in either type IIB or type IIA supergravity, yet we have insisted on type IIB theory. This preference can be attributed to the fluxes, which will only scale correctly in type IIB supergravity, suggesting that the KK reduction carries some memory that it was originally a reduction of type IIB supergravity on S$^5$.

One could first take a limit where the S$^5$ degenerates to S$^3\times \R^2$ \cite{Cvetic:1999pu} and then apply our scaling limit. This enables our KK reduction to be reinterpreted as arising from a scaling limit of an S$^3$ reduction of type IIA theory \cite{Cvetic:2000ah} followed by a toroidal reduction, though the lower-dimensional theory would have a domain wall as a vacuum solution rather than (A)dS.

We have omitted natural three-form fluxes, both NS-NS and RR, in our Ansatz. Though these scale correctly with the warp factor, we find that it is difficult to include them in an Ansatz, since they would require a decomposition into an external two-form and internal one-form. For the internal space $\mathbb{R}^4$, this is inconsistent with internal Ricci-flatness, and for Calabi-Yau reductions one generically does not have natural internal one-forms. It is expected that when $\lambda =0$, modulo mirror symmetry (T-duality), we recover the Ansatz of the reduction of type IIB supergravity on CY$_2$, e. g.  \cite{Duff:1998cr}.

\subsection{KK reductions on $\mathbb{R}^3$}
Switching our attention to the final warped-product solution, which lives in eleven dimensions, we can construct a KK reduction Ansatz with the internal odd-dimensional space further split as $\mathbb{R}^3 \equiv \mathbb{R}^2 \times \mathbb{R}$.

The Ansatz is motivated by a scaling limit of $D=7$ gauged supergravity and is given as
\bea
\dd s_{11}^2 &=& \Delta^{\frac{2}{3}} \dd s^2(M_8) + \Delta^{-\frac{4}{3}}L^2 \biggl[ e^{-2Y} \dd \mu_0^2 \nn && \phantom{xxxxxxxxxxxxxxxx}+ e^{Y} ( \dd \mu_2^2 + \mu_2^2 \dd \psi_2^2) \biggr], \nn
G_4 &=& H \wedge \dd \mu_0,
\eea
where $\Delta = \sqrt{1 + \lambda( \mu_0^2 + \mu_2^2)}$ and $H = \dd B_2$. The Ansatz for the four-form flux is largely self-selecting in that it preserves the symmetry of the $\mathbb{R}^2$ factor and scales correctly with the warp factor $\Delta$ in the Einstein equation.
We remark that this choice appears to fall outside the Ansatz of Cvetic et al \cite{Cvetic:1999xp} but can be accommodated in reductions on S$^4$, where three-forms are retained \cite{Liu:1999ai, Nastase:1999cb}.

After the KK reduction, the eight-dimensional action is
\bea
\label{D8_theory}
\mathcal{L}_8 = \sqrt{-g_8} \biggl[ R_8 - \frac{3}{2} (\partial Y)^2 &-& \frac{1}{12} e^{2Y} H_{abc} H^{abc} \nn &-& \frac{2}{L^2} \lambda (e^{2 Y} + 2 e^{-Y} )\biggr].
\eea

Note that one can also find a KK reduction Ansatz in type IIA theory by first taking a limit where the S$^4$ degenerates to S$^3\times \R$ \cite{Cvetic:1999pu,Cvetic:2000ah}, reducing along the $\R$ direction and then performing our scaling limit.

\section{ Stability of AdS vacua}
\label{sec:stability}
It is a well-appreciated fact that in the absence of supersymmetry classical stability is a concern. In this section we focus on the stability of the (A)dS vacua. We begin with AdS stability, where violations of the Breitenlohner-Freedman (BF) bound \cite{BF-paper} provide us with a simple litmus test for instability.
We will turn our attention to dS vacua later.

An important caveat from the outset is that we will confine our attention to instabilities that arise \textit{within} our truncation Ansatz. However, this does not preclude the possibility that instabilities arise from modes we have truncated out, for example see \cite{Bobev:2010ib}. Within this restricted scope, we will explicitly show that lower-dimensional Freund-Rubin type solutions with AdS$_3$ factors enjoy greater stability than higher-dimensional vacua in the truncated theory. This may be attributed to the fact that these solutions correspond to the near-horizon of known black holes and are expected to be classically stable.

That some of our geometries are unstable may come as no surprise. It is known that product spaces can be prone to instabilities where one space becomes uniformly larger while another shrinks so that the overall volume is kept fixed. Earlier examples of this instability include the spacetimes AdS$_4 \times$ M$_n \times$ M$_{7-n}$ \cite{Duff:1984sv} and AdS$_7 \times$ S$^2 \times$ S$^2$ \cite{Berkooz:1998qp}. Indeed, in a fairly comprehensive study of the classical stability of Freund-Rubin spacetimes of the form AdS$_p \times$ S$^q$ \cite{DeWolfe:2001nz}, this is the primary instability observed.

In terms of our Ansatz, the scalars in the lower-dimensional theory control the volume of the internal spaces, which are further subject to a constraint. We will now investigate the stability of geometries with respect to these scalar modes, while at the same time taking into account breathing modes that arise from further reductions. Our analysis is not intended to be comprehensive but, since instabilities usually arise from products \cite{DeWolfe:2001nz}, experience suggests these are the most dangerous modes. We recall that the BF bound for AdS$_p$ with radius $R_{\textrm{AdS}}$ is \cite{BF-paper}
\be
\label{BF}
m^2 R_{\textrm{AdS}}^2 \geq - \frac{(p-1)^2}{4}.
\ee

Before proceeding to specialize to particular cases, here we record some preliminaries. We will in general consider spacetimes in dimension $D$ and further reductions on constant curvature spaces of dimension $(D-p)$ to a $p$-dimensional spacetime:
\be
\dd s^2_{D} =  e^{2A \frac{(D-p)}{(2-p)}} \dd s^2_{p} + e^{2A} \dd s^2 (\Sigma_{D-p}),
\ee
which leads to scalars $A$, i.e. breathing modes, in the lower-dimensional theory.

The above reduction Ansatz is designed to bring us to the Einstein frame in $p$ dimensions. The Einstein-Hilbert term in the higher-dimensional action reduces to
\bea
\label{EH_term}
\mathcal{L}_p &=& \sqrt{-g_p} \biggl[ R - \frac{(D-2)(D-p)}{(p-2)} (\partial A)^2 \nn && \phantom{xxxx} + e^{- \frac{2(D-2)}{(p-2)} A} \kappa (D-p-1) (D-p) \biggr],
\eea
where $\kappa$ is the curvature of the internal space. When $\kappa >0$, we will only consider the constant spherical harmonic, which appears with the lowest mass, $\nabla^2_{S^{q}} A = 0$. Higher spherical harmonics typically do not lead to instabilities, since they correspond to modes with a more positive mass squared. We can now specialize to the various potentials we have found.

\subsection{$D=11, p=5,\ \Sigma_{D-p}=\mathbb{R}^6$}
Since our lower-dimensional theory corresponds to the bosonic sector of a known supergravity, it is expected that solutions are stable.
It is easy enough to check that the scalars precisely saturate the (unit radius) BF bound in five dimensions:
\be
\nabla_{AdS_5}^2 \delta \varphi_i + 4 \delta \varphi_i = 0.
\ee

At this point, it is also instructive to make a comment regarding reductions in the \textit{absence} of fluxes. The first non-trivial reduction with an AdS$_3$ vacuum involves a reduction on $H^2$. We omit the details since a related reduction appeared recently in \cite{Karndumri:2013iqa} where, in addition, the underlying three-dimensional gauged supergravity was identified. The mass-squared matrix for the fluctuations takes the following form:
\be
\nabla^2_{AdS_3} \left( \begin{array}{c} \delta \varphi_1 \\ \delta \varphi_2 \\ \delta A \end{array} \right) = \left( \begin{array}{ccc} -2 & 0 & 0 \\ 0 & -2 & 0 \\ 0 & 0 & 4 \end{array} \right) \left( \begin{array}{c} \delta \varphi_1 \\ \delta \varphi_2 \\ \delta A \end{array} \right).
\ee
We observe that the scalars $\varphi_i$ have mass $m^2 = -2$ which violates the BF bound for AdS$_3$. In contrast, the fluctuation of the scalar $A$ does not affect the stability. Indeed, it is worth observing that we can also truncate out $\delta \varphi_i$, in which case the instability would not be observed. This is a pretty trivial example of the hidden instabilities noted in a higher-dimensional context in \cite{Bobev:2010ib}.

\subsection{$D=10, p=6,\ \Sigma_{D-p}=\mathbb{R}^4$}

In contrast to the AdS$_5$ vacuum, the AdS$_6$ vacuum is unstable to fluctuations of the scalar $Z$ with a mass squared $m^2 = -10$ for unit-radius AdS$_6$.  We will now consider whether a lower-dimensional AdS$_3 \times \Sigma_3$ vacuum is stable to this mode. Projecting out the massless axion and the dilaton, which are less likely to source instabilities, the theory may be dimensionally reduced to give
\bea
\mathcal{L}_3 &=& \sqrt{-g_3} \biggl( R_3 - \frac{1}{4} (\partial Z)^2 - 12 (\partial A)^2 + 6 \kappa e^{-8A} \nn &+& \frac{8}{L^2} e^{-6A} \cosh \frac{Z}{2} - \frac{1}{2}  e^{-12A} \left[ p^2 e^{Z}+ q^2  e^{ - Z}\right] \biggr), \nonumber
\eea
where we have assumed
\be
H_3 = p \, e^{Z-12 A} \vol(M_3) + q \vol(\Sigma_3).
\ee
In the electric term in $H_3$, we have imposed the lower-dimensional equation of motion.  Extremizing the potential, we get
\bea
p^2 &=& 4 e^{4A-Z}\left( \kappa + L^{-2} \, e^{\frac{1}{2} Z+2A} \right) ,\nn q^2 &=& 4 e^{4A+Z}\left( \kappa + L^{-2} \, e^{-\frac{1}{2} Z+2A} \right).
\eea
Plugging these back into the action, we note that AdS$_3$ will have unit radius provided that
\be
\label{unit_radius}
\frac{4}{L^2} \cosh \frac{Z}{2} = 2 e^{6A} - 2 \kappa e^{-2A}.
\ee
The AdS$_3 \times$S$^3$ solutions appeared previously in \cite{Fareghbal:2008ar}.

If one considers variations of the breathing mode $A$ and scalar $Z$, then the mass-squared matrix has eigenvalues
\bea
\label{D6_mass}
m^2 &=& \frac{4e^{-16A}}{L^4} \biggl( 2 e^{8A} L^4 \kappa + 2 e^{10 A} L^2 \cosh \frac{Z}{2}\nn  &\pm& \sqrt{e^{20 A} L^4 (2 \cosh Z -1)} \biggr).
\eea
To simplify expressions, we can solve (\ref{unit_radius}) for the length scale $L$ in terms of $\kappa, A$ and $Z$:
\be
L = \sqrt{2} e^{-A} \sqrt{\frac{\cosh \frac{Z}{2}}{e^{4A}-\kappa e^{-4A}}}.
\ee
Since (\ref{D6_mass}) is symmetric under $Z \leftrightarrow -Z$, without loss of generality we can take $Z \geq 0$. For both $\kappa = 0$ (T$^3$) and $\kappa =1$ (S$^3$), we find that the masses are always strictly positive. Indeed, for $\kappa =0$, all dependence on the critical value of $A$ drops out and $m^2 \rightarrow 0^{+}$ as $Z \rightarrow  \infty$ for the lowest eigenvalue.  When $\kappa=1$, the dependence on $A$ remains, but $m^2$ is again strictly positive.

When $\kappa = -1$, we have a constraint on the range of $A$ and $Z$, namely $A \geq \frac{z}{8}$,  such that the solution is real, i.e. $p^2, q^2 \geq 0$. In this range, $m^2$ is always positive. Thus, we conclude that a Freund-Rubin type AdS$_3$ solution is stable to fluctuations that destabilize the AdS$_6$ vacuum of the six-dimensional theory.

\subsection{$D=11, p=8,\ \Sigma_{D-p}=\mathbb{R}^3$}
The stability analysis for the eight-dimensional theory \eqref{D8_theory} parallels the six-dimensional theory \eqref{D6_theory}, which we analyzed previously. Even in the absence of supersymmetry, where there is no dual (super)conformal theory in seven dimensions \cite{Nahm:1977tg}, the AdS$_8$ vacuum is puzzling, but it resolves itself by being unstable. In order to stabilize the vacuum, one can turn on the three-form. One can support a ``magnetovac" solution, AdS$_5 \times \Sigma_3$, however the fluctuations of the scalars $A$ (breathing) and $Y$ have the following mass eigenvalues (at unit radius):
\be
m^2 = 4 (3 - 4 e^{-3 Y} \pm \sqrt{25 -32 e^{-3Y} +16  e^{-6 Y}}\ ).
\ee
and are thus unstable for all $\Sigma_3$.

We can also consider an ``electrovac" with an AdS$_3$ factor but this solution can be incorporated in a more general case:
\bea
\dd s^2 &=& e^{-2A} \left[ e^{-4B} \dd s^2(M_3) + e^{2B} \dd s^2(\Sigma_2) \right] + e^{2A} \dd s^2(\Sigma_3), \nn
H &=& p e^{-2Y -4A -8B} \vol(M_3) + q \vol(\Sigma_3),
\eea
where we now have two breathing modes $A$ and $B$, two Freund-Rubin-type flux terms $p, q$ and a transverse space $\Sigma_2$ of constant curvature $\kappa_2$. We can view the Ansatz as two successive reductions, one on $\Sigma_3$, followed by a second on $\Sigma_2$, where in each case one arrives in the Einstein frame.

The effective three-dimensional action may be written as
\bea
\mathcal{L}_3 &=& \sqrt{-g_3} \biggl( R_3 - \frac{3}{2} (\partial Y)^2 - 6 (\partial A)^2 -6 (\partial B)^2 \nn &+& \frac{2}{L^2} (e^{2Y} + 2 e^{-Y}) e^{-2A-4B} - \frac{1}{2} p^2 e^{-2Y-4A -8B} \nn
&+& 6 \kappa_1 e^{-4A-4B} + 2 \kappa_2 e^{-6B} - \frac{1}{2} q^2 e^{2Y-8A-4B}\biggr)
\eea
where $\kappa_1$ denotes the constant curvature of $\Sigma_3$. Extremizing the potential, we arrive at the critical point:
\bea
\label{crit_point}
e^{2Y} &=& - L^2 \kappa_2 e^{2A-2B}, \nn
q^2 &=& 4 \kappa_1 e^{-2Y+4A} + \frac{2}{L^2} e^{6A}, \\
p^2 &=& 4 \kappa_1 e^{2Y+4B} + \frac{4}{L^2} (e^{-Y}-\frac{1}{2} e^{2Y}) e^{2Y+2A+4B}. \nonumber
\eea
We observe that the Riemann surface $\Sigma_2$ should be negatively curved, thus making it $H^2$.
We can set $q=0$, $\frac{1}{2} \kappa_1 = \frac{1}{4} \kappa_2 = \kappa$ and $B=2A$ to recover the electrovac solution, the details of which we have omitted. We can set AdS$_3$ to unit radius by choosing
\be
L^2 = \frac{e^{-Y+2A}}{(e^{4A+4B}-\kappa_1)}.
\ee

\begin{figure}[h]
\label{k0}
\centering
\includegraphics[width=0.4\textwidth]{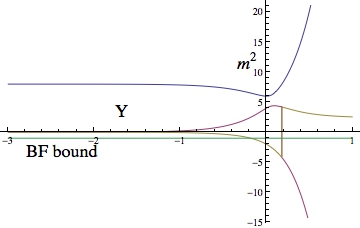}
\caption{The mass-squared eigenvalues for scalar fluctuations around the geometry AdS$_3 \times$ T$^3 \times$ H$^2$ as a function of the critical value of $Y$.}
\end{figure}

With an additional breathing mode, the stability analysis is more complicated. In fact, even for the simpler case when $\kappa_1 =0$, where expressions do not depend on the breathing modes $A$ and $B$, we cannot find analytic expressions for the mass-squared matrix eigenvalues. For $\kappa_1 =0$, we note that 
in order the solution to remain real, $Y$ is restricted to the range $Y \leq \frac{1}{3} \log 2$. The mass as a function of the critical value of $Y$ is plotted in FIG. 1. We see that for suitably negative values of $Y$, a range of stability exists. As can be seen from (\ref{crit_point}), this corresponds to values where the fluxes are larger. 

When $\kappa_1 = \pm 1$, the mass-squared matrix only depends on $Y$ and the combination $A+B$. It is easier to consider $\kappa_1 = -1$, since we have a constraint. From $p^2, q^2 \geq 0$, we find a constraint on $Y$ in terms of $A+B$:
\bea
\label{k1_minus}
\frac{1}{2} (e^{4 A+4 B}+1) &\geq& e^{-3Y} \geq \frac{1}{2} \frac{(e^{4 A+4 B}+1)}{e^{4A+4B}}.
\eea
Taking $A+B$ to be a fixed value, for either value of $\kappa_1$, we see that the eigenvalues of the mass matrix vary with the critical value of $Y$ in essentially the same way as they do for the AdS$_3 \times$ T$^3 \times$ H$^2$ geometry shown in FIG. 1. Thus, the eigenvalues of the mass-squared matrix are largely insensitive to curvature, given our choice of normalization. This means that the same range of stability will exist for $\kappa_1 = \pm 1$. The only caveat here is that for $\kappa_1 = -1$, there is the added constraint above (\ref{k1_minus}), so $A+B$ has to be chosen to be large enough such that one has some overlap with the stable region.

\section{de Sitter vacua}
In this section we discuss a particular dS$_3$ solution in the five-dimensional theory \eqref{U13theory}, which via our consistent KK reduction may be regarded as a solution to eleven-dimensional supergravity.  Neglecting time-dependent solutions, such as \cite{Townsend:2003fx}, static embeddings in the literature have either involved reductions on non-compact hyperbolic spaces, for example \cite{Hull:1988jw, Gibbons:2001wy, Lu:2003dm}, or analytic continuations of known maximally supersymmetric solutions, such as AdS$_5 \times$ S$^5$, leading to solutions of so-called type II$^{*}$ theories  and their dimensional reductions \cite{Hull:1998vg,Lu:2003dn}.  Here we point out that the internal non-compact spaces need not be curved and can in fact  be Ricci-flat. This evades the well-known ``no-go" theorem \cite{Maldacena:2000mw} on non-compactness grounds.

Our effective three-dimensional theory supporting the dS$_3$, comes from a reduction of the U(1)$^3$ theory \eqref{U13theory} on a Riemann surface of constant curvature $\kappa$ \footnote{It should be noted that the original five-dimensional vacuum corresponds to a local maximum and is unstable.}. To do this, we employ the usual Ansatz,
\bea
\dd s^2_5 &=& e^{-4A} \dd s^2_3 + e^{2A} \dd s^2(\Sigma_2), \nn
F^i &=& - a_i \vol(\Sigma_2),
\eea
where $a_i$ denote constants. The three-dimensional action may be recast as
\bea
\mathcal{L}_3 &=& \sqrt{-g_3} \left[ R_3 - \frac{1}{2} \sum_{i=1}^3 (\partial W_i)^2 - V(W_i) \right]
\eea
where the potential $V$ takes the form
\bea
\label{pot}
V &=& -2 \kappa e^{K} +  \frac{4}{L^2} e^{K} \sum_{i=1}^3  e^{W_i} + \frac{1}{2} e^{2K} \sum_{i=1}^3 a_i^2 e^{2 W_i}.
\eea
In expressing terms this way, we have made use of the K\"{a}hler potential of the three-dimensional gauged supergravity, $K = -(W_1 + W_2 +W_3)$, introduced a length scale $L$ for the internal space, and imported the notation of \cite{Karndumri:2013iqa}, $e^{W_i} = e^{2A} X_i^{-1}$, where $A$ denotes the warp factor.

Up to the minus sign in front of the second term, this is the potential corresponding to magnetized wrapped brane solutions \cite{Karndumri:2013iqa}. This potential has an underlying real superpotential provided $\kappa = - (a_1 + a_2 + a_3)$. One advantage of working with the type II$^{*}$ embeddings is that flux terms appear with the ``wrong" sign and the theories may be regarded as ``supersymmetric". Solutions then follow from extremizing the fake superpotential. This is not the case here, since the flux terms do not have the wrong sign. We have checked that a fake superpotential can be found, but only when all the constants are equal, $a_i = a$,  and $\kappa = 5 \frac{a}{L}$. One of the extrema of the potential in this case is AdS$_3$, so we will ignore this possibility.

Extremizing \eqref{pot}, we arrive at conditions on the fluxes for a critical point to exist:
\be
\label{dS_ai}
a_i^2 = e^{\sum_{j\neq i} W_j} \left( \kappa e^{-W_i} -\frac{4}{L^2}\right).
\ee
For real solutions we recognize the immediate need for a reduction on a sphere $(\kappa > 0)$. Inverting the above expression to get $W_i$ in terms of $a_i$ is, in general, problematic, so we consider the simplification where $a_i = a$, $W_i = W$. In this case, it is easy to locate the critical points of $V$,
\be
e^{W_{\pm}} = \frac{L^2}{8} \left( 1 \pm \sqrt{1 - 16 a^2 L^{-2} }\right).
\ee
We note that we require $16 a^2 < L^2$ for two real extrema. Examining the second derivative of the potential, we identify the upper sign as a local maximum and the lower sign as a local minimum corresponding to our de Sitter vacuum. By tuning the parameter $a$ relative to $L$, as we show in FIG. 2, it is possible to find a de Sitter vacuum, where the cosmological constant is arbitrarily small and positive.

\begin{figure}[h]
\label{desitter}
\centering
\includegraphics[width=0.4\textwidth]{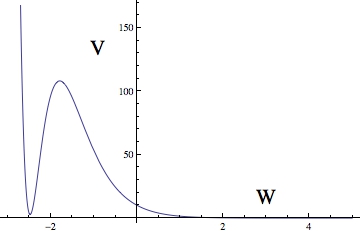}
\caption{Plot of the potential for $L=1$ and $a = 0.236$. By tuning $a$ relative to $L$, we can increase the barrier to decay and stabilise the vacuum.}
\end{figure}

To address stability, we follow the treatment presented in \cite{KKLT}, which is based in part on \cite{Coleman:1980aw}. Since we are working in $D=3$, it is natural to consider an O(3)-invariant Euclidean spacetime with the metric,
\be
\dd s^2 = \dd \tau^2 + a(\tau)^2 \dd \Omega_2^2,
\ee
where $a$ is the Euclidean scale factor. The scalars obey the following equations of motion,
\be
3 W'' + 6 \frac{a'}{a} W' = V,_{W}, ~~a'' = - \frac{a}{2} \left( \frac{3}{2} W'^2+V \right),
\ee
where primes denote derivatives with respect to $\tau$. These equations admit a simple instantonic three-sphere solution, where the scalar sits at one of the extrema of the potential, $W=W_{\pm}$, and
\be
a(\tau) = \ell^{-1} \sin( \ell \tau).
\ee
Here $\ell$ is the inverse radius of the sphere, which in turn is related to the potential, $ \ell^2 = \frac{V}{2}$. Given the two extrema, we have two trivial solutions of this type describing a time-independent field.

We now wish to consider Coleman-De Luccia instantons, which describe tunneling trajectories between the de Sitter vacuum and asymptotic Minkowski space ($W > W_+$). According to \cite{Coleman:1980aw}, the probability, $P$, for tunneling from from a false vacuum at $W = W_-$, with vacuum energy $V_0 \ll 1$ (in Planck units), to Minkowski space is to first approximation given by
\be
P \approx  \exp(S_0),
\ee
where $S_0 \equiv S(W_0)$ is the Euclidean action evaluated in the vicinity of the de Sitter vacuum. $S_0$, in turn, is determined from the tunneling action,
\be
\label{Euclid_action}
S(W) = \int d^3 x \sqrt{g} \left( -R_3 +\frac{3}{2} (\partial W)^2 + V(W) \right),
\ee
which describes trajectories beginning in the vicinity of the false vacuum, $W=W_-$, at $\tau =0 $ and reaching $W =0$ (Minkowski) at $\tau = \tau_f$, where $a(\tau_f) = 0$. Using the trace of the Einstein equation, we can rewrite (\ref{Euclid_action})
\be
S(W) = - 2 \int d^3 x \sqrt{g} V(W)= - 8 \pi \int_0^{\tau_f} d \tau a^2(\tau) V(W(\tau)).
\ee
The Euclidean action calculated for the false vacuum de Sitter solution at $W= W_{-}$ is
\be
S_0 = - 8 \pi^2 \sqrt{\frac{2}{V_0}}.
\ee
By tuning $a$ and $L$ appropriately, so that $V_0$ is small, we can find an arbitrarily long-lived dS$_3$ vacuum. We conclude that the dS$_3$ vacuum can be regarded as stable. Though we have only presented one example, we expect similar comments to hold for dS$_3$ vacua supported through the consistent reductions we have identified.

As it stands, our set-up needs some tweaking in order to incorporate dS$_4$ vacua. We have seen that a vacuum solution exists when the number of internal dimensions is large. In the absence of two-form flux in the six-dimensional theory \eqref{D6_theory}, one could contemplate reducing on a 2d Riemann surface. For the dS$_4 \times$ S$^2$ solution without flux threading the S$^2$, it is not surprising that one finds that the vacuum is unstable.

Before leaving the subject of de Sitter solutions, we make one final comment. The five-dimensional theory (\ref{U13theory}) also has solutions that smoothly interpolate between dS$_2\times$ S$^3$ in the infinite past and a dS$_5$-type spacetime in the infinite future \cite{Lu:2003iv}. These solutions can be obtained from the AdS black hole solutions in $D=5$ U(1)$^3$ gauged supergravity simply by changing the sign of the scalar potential, and can all be embedded in eleven dimensions using the KK reduction Ansatz (\ref{redansatz}). Like the previously-mentioned dS$_3$, these solutions have a fake superpotential in the equal-charge case.

\section{Discussion}

We have studied a class of non-supersymmetric Ricci-flat solutions which are warped products of a flat internal space and an anti-de Sitter or de Sitter spacetime.  We have found that these Ricci-flat solutions are limited to three cases: warped products of (A)dS$_5$ and $\R^6$ in eleven dimensions, (A)dS$_8$ and $\R^3$ in eleven dimensions, and (A)dS$_6$ and $\R^4$ in ten dimensions. There is also a fourth potentially interesting case of (A)dS$_4$ in a spacetime with large dimension $D$. Given that these geometries are rather simple and do not involve any matter content, it is intriguing that so few examples exist and that they are mainly limited to ten and eleven dimensions. While singular in the anti-de Sitter cases, these geometries are completely smooth for de Sitter and are similar in structure to the ``bubble of nothing'' \cite{Bubble-nothing}.
Unlike direct products of AdS and a sphere or warped products of de Sitter and a hyperbolic space, both of which are supported by flux, our solutions are supported entirely by the warp factor, are hence not bound by the no-go theorem \cite{Maldacena:2000mw}, and do not appear to arise from a limit of the former solutions.

We construct consistent KK truncations for which the above solutions arise as vacuum solutions. These KK truncations are shown to arise as limits of the celebrated dimensional reductions on spheres (like those discussed in \cite{Cvetic:1999xp}) in which the lower-dimensional spacetime gets augmented by one of the spherical coordinates while the remaining directions along the sphere get flattened out. Unlike KK truncations on hyperbolic spaces which are associated with non-compact gauge groups, the truncations in this paper lead to the bosonic sector of gauged supergravities with compact gauge groups. This is because the gauge fields are associated only with the flux in the higher-dimensional theory and, rather surprisingly, not the geometry. Therefore, within this truncation the isometries of the internal space do not play an explicit role in the lower-dimensional (bosonic sector of gauged) supergravity. This KK reduction enables one to embed five-dimensional U(1)$^3$ de Sitter gravity in eleven-dimensional supergravity.  It is an interesting open direction to consider generalizations where the gauge groups are non-Abelian. We expect that one can achieve this by considering a similar limit of the maximally supersymmetric SO(4) \cite{Nastase:1999cb}, SO(6) \cite{Cvetic:2000nc}
 and SO(8) reductions \cite{s7}.

Given that our solutions do not preserve supersymmetry, it is important to study their stability. We have focused on possible classical instabilities associated with breathing modes, though there could be other instabilities associated with massive modes that have been truncated out. Within this limited setting, one finds that the AdS$_5$ solution, although corresponding $D=11$ solution is singular, is stable. A dual four-dimensional non-supersymmetric CFT is a rather intriguing notion, given that this is dual to pure $D=11$ gravity (reduced on our Ricci-flat solutions), that D-branes are completely absent and that the eleven-dimensional solution is singular. Other stable solutions include AdS$_3\times \Sigma_3$, where $\Sigma_3$ is S$^3$, T$^3$ or H$^3$, as well as AdS$_3\times \Sigma_3 \times$ H$^2$ for a certain range of its parameters. On the other hand, the AdS$_8$, AdS$_6$ and AdS$_5\times \Sigma_3$ solutions are not stable. It would have been rather surprising if there had been a stable AdS$_8$ solution, since this would imply the existence of a corresponding seven-dimensional CFT, though it would be non-supersymmetric.

As for the de Sitter solutions, we find dS$_3\times$ S$^2$ solutions that, in terms of the breathing modes, are stable. We expect similar solutions to of the form dS$_3 \times$ S$^3$ and dS$_3 \times$ S$^3 \times$ H$^2$ to exist in the six-dimensional \eqref{D6_theory} and eight-dimensional theory \eqref{D8_theory}, respectively. Although all the dS$_4$ vacua we have found are either i) unstable or ii) they require an infinite number of internal dimensions, and are thus unsatisfactory, we hope that this line of inquiry will lead to simple stable dS$_4$ in the future. The most positive angle is that a flat direction in the reduction on $\mathbb{R}^6$ to $D=5$ can be found and a dS$_4$ vacuum can be engineered from the dS$_5$ vacuum using the approach of  \cite{Klebanov:2004ya}.

As with warped-product solutions of de Sitter and a hyperbolic space \cite{Gibbons:2001wy}, our solutions appear to be intrinsically higher dimensional, in that they are not amenable to either compactification or a braneworld scenario \cite{RS}. In particular, massless gravitational modes are not associated with normalizable wavefunctions and therefore cannot be localized on a brane. It is not clear as to whether one can use the proposed dS/CFT correspondence \cite{Strominger} to extract meaningful information directly from the eleven-dimensional embeddings of these de Sitter solutions.  

While our construction remains intact if the internal space $\mathbb{R}^n$ is replaced by a cone over any Einstein space with positive curvature, there is a conical singularity at the apex of the cone. Replacing the internal space with a smooth cone, such as a resolved or deformed conifold \cite{Candelas} for the case of a six-dimensional space, would be interesting but would necessitate a slightly different Ansatz than was considered in this paper.

\section*{Acknowledgements}	
We have enjoyed fruitful discussions with K. Balasubramanian, C. Hull, K. Jensen, J. Liu, H. Lu, P. Meessen, D. S. Park, C. N. Pope, M. Ro\v{c}ek, J. Schmude, K. Stelle, P. van Nieuwenhuizen \& O. Varela. The work of E. \'O C is supported by Marie Sklodowska-Curie grant 328625 ``T-Dualities".

\end{document}